# Replicating and Applying a Neuro-Cognitive Experimental Technique in HCI Research

**David Coyle**

Interaction and Graphics Group,

Dept. of Computer Science,

University of Bristol,

Bristol BS8 1UB, UK

david.coyle@bristol.ac.uk

## Abstract

In cognitive neuroscience the sense of agency is defined as the as the experience of controlling one's own actions and, through this control, affecting the external world. At CHI 2012 I presented a paper entitled "I did that! Measuring Users' Experience of Agency in their own Actions" [1]. This extended abstract draws heavily on that paper, which described an implicit measure called *intentional binding*. This measure, developed by researchers in cognitive neuroscience, has been shown to provide a robust implicit measure for the sense of agency. My interest in intentional binding stemmed from prior HCI literature, (e.g. the work of Shneiderman) which emphasises the importance of the sense of control in human-computer interactions. The key question behind the CHI 2012 paper was: can we apply intention binding to provide an implicit measure for the experience of control in human-computer interactions? In investigating this question, replication was a key element of the experimental process.

## ACM Classification Keywords
H.5.2 [Information Interfaces and Presentation]: User Interfaces - Evaluation/methodology;

**Intentional Binding**

Repeated experiments have shown that voluntary human actions are associated with systematic changes in our perception of time [3]. The interval between a voluntary or intentional action and the outcome of such an actions is typically perceived as shorter than the actual interval. For example, if a person voluntarily presses a button and this action causes an outcome - e.g. a beep - it is highly likely that the person will perceive their action as having happened later than they it actually did (action binding). They are also likely to perceive the outcome as having happened earlier than it actually did (outcome binding). Patrick Haggard, the research who first identified this phenomenon, coined the term *'Intentional Binding'* to describe it, as it is contingent on several factors [3]. In the absence of outcomes people are found to more accurately report the timing of actions. For the temporal binding effect to occur, actions must be *intentional* and must lead to an *outcome*. Under these conditions our perception of the timings of actions and their outcomes become bound together temporally.

In the years since Haggard's first experiments, a large number of studies have validated and built on his initial observations. In and of itself this repeated experimentation highlights the importance of replication in cognitive neuroscience research. Based on this replication, a scientific consensus is now supports the conclusion that time perception in voluntary actions - and the binding effects associated with such actions - provides a robust implicit metric for the sense of agency. Higher intentional binding values correlate to a greater sense of personal agency.

**Replication and application**

Detailed descriptions of the experimental methods used to assess intentional binding are beyond the scope of this short paper. These details are available in the CHI 2012 paper [1]. Instead I will focus more broadly on the ways in which we replicated prior experiments and applied this metric.

*Experiment 1*

Neuro-cognitive experiments on intention binding typically focus on very simple interactions, e.g. a button press that causes a beep. My first experiment focused on the modality of the interaction. It asked if changes in the modality of an interaction lead to changes in the sense of agency. The experimental design closely mirrored procedures originally outlined by Haggard. One independent variable was manipulated: the input modality. We compared a traditional input device - a keypad - with a skin-based input device. The keypad replicated the input typically used in neuroscience research. In condition one the participant pressed a button on a keypad to cause a beep. In condition two - the skin-based condition - the participant caused a beep by tapping on their arm. The skin-based capture device was attached to the participant's left arm and they tapped this arm with their right hand. In all cases there was a fixed interval of 250ms between the participant's action and the beep.

Results showed that users experienced significantly higher intentional binding for skin-based interactions. Across 19 participants a mean binding of 42.92ms was observed in the button press condition. I.e. an interval of 250ms was perceived as 207.08ms. Importantly, this binding value is consistent with the results of prior binding experiments that have used button inputs. In the skin-based condition participants experienced a total binding effect of 109.47ms. Here 250ms was perceived as 140.53ms. Given the correlation between intentional binding and the sense of agency, this experiment suggests that people experience a significantly greater sense of agency, or control, when they interact with technology via skin-based input, as compared with traditional keypad input.

More broadly speaking, this experiment provided empirical evidence that different interaction modalities can provide different experiences of control and ownership. In undertaking this experiment I believe it was essential that one of our input conditions - the keypad - replicated prior cognitive neuroscience research. This replication demonstrates that our experiment was administered effectively and lends strength and credibility to our findings. It also allows our results to be judged against and incorporated into the prior body neuro-cognitive research on intentional binding and the sense of agency.

Ultimately I hope the method we introduced can be used to investigate the sense of agency across a wide range of input modalities. For other researchers using this technique, I strongly recommend that replication (plus extension) of prior results again be a key element in the design of new experiments.

*Experiment 2*
Cognitive neuroscience experiments on intentional binding have typically examined voluntary and involuntary actions. From an HCI perspective, this might be considered an unnecessarily black or white disjunction. Many user interactions with technology are more intermediate. In particular 'intelligent' user interfaces often seek to interpret and act on the intentions of the user. Here users' actions are voluntary, but the outcomes may be assisted. The second experiment in the CHI 2012 paper was designed to investigate users' sense of agency in interactions where a computer interprets their intention and helps them to achieve a goal. In this sense the second experiment diverged further for the interactions examined in prior cognitive neuroscience research. However, as in first experiment, we apply an experimental procedure that closely matched prior literature.

The experiment investigated agency in a machine-assisted point-and-click task. Using a mouse, participants were required to hit targets on a computer screen, as quickly and as accurately as possible. The computer provided assistance through an algorithm that effectively added gravity to targets, thereby making it easier for participants to complete the task. Hitting a target caused a beep. In each trial there was a random interval between hitting a target and the beep, and participants were asked to estimate this interval.

In the experiment we investigated four different assistance levels, which varied from no assistance to a very high, and very obvious, level of computer assistance. Results suggested that, up to a certain point, the computer could assist users whilst also allowing them to retain a sense of agency for their actions. However, we found that beyond a certain level of assistance users experienced a detectable loss in their sense of agency. This loss in agency occurred in spite of the fact that the computer correctly interpreted users' intentions and assisted them in achieving their goal.

Our results suggest that for the assisted input algorithm we investigated - and possibly for assisted input systems more generally - there may exist a tipping point or sweet spot. This is the point at which a computer can help people and potentially maximise task performance - e.g. speed or accuracy - without significant detriment to the experience of agency. I find this possibility very intriguing. However I also believe further investigation, and further replication, is required to assess the generalizability of our initial finding. I am currently undertaking such research.

## Conclusions

Alongside the issues discussed above, I have one minor comment on the CHI submission process. When I submitted the original CHI 2012 paper, I was very keen to also submit the dataset for my studies. Under the 2012 submission system this was not possible. I understand that this issue was addressed for CHI 2013 submissions. This was a real step forward.

## Citations